\documentclass[12pt]{article}
\usepackage{amssymb}

\usepackage{latexsym}
\usepackage{graphicx}
\usepackage{hyperref}

\graphicspath{{./}{E:/SWP30/MyDocs/}}
\begin{document}

\title{Linearly Coupled Directed Percolation in the Strong Coupling Regime}
\author{R. Dengler \\
\textit{Rohde }\&\textit{\ Schwarz GmbH }\&\textit{\ Co KG, 1GP1}\\
\textit{M\"{u}hldorfstr. 15, 81671 Munich, P.O.B. 801469.}}
\maketitle

\begin{abstract}
We consider directed percolation processes for particle types $A$ and $B$
coupled unidirectionally by a transmutation reaction $A\rightarrow B$. It is
shown that the strong coupling regime of this recently introduced problem
defines a universality class with upper critical dimension $d_{c}=6$. Exact
expressions are derived for the scaling dimensions in the inactive phase
above $d=4$. Below $d=4$ the interactions of the normal directed percolation
also get relevant.\medskip \smallskip

\noindent PACS numbers: 64.60.Ak, 64.60.Ht,64.60.Kw.

\smallskip
\end{abstract}

It is a well known fact that the phase diagram (or more generally, the
parameter space) of homogeneous media may contain critical submanifolds.
Near such a critical manifold some local restoring forces vanish and the
system becomes ``soft'' with respect to some external forces. As a
consequence next-neighbour interactions and long-range correlations begin to
play an important role. If in addition a random force is acting in the
system, then a complex behaviour may result. The most important experimental
aspects of critical phenomena are scale invariance and universality.

The universality classes may be subdivided into categories like ``Critical
Statics'', ``Critical Dynamics'', ``Reaction-Diffusion'' and
``Self-Organized''. Dozens of universality classes have been identified.
Formally there are infinitely many universality classes - most of them
without a plausible physical interpretation, of course.

The \emph{directed percolation} (DP) universality class{}\cite{Obu80}\cite
{CS80} is one of the simplest and most important in the reaction-diffusion
type category. It occurs in contexts like birth-death processes, population
growth and chemical reactions.{}\cite{Sch72} It also may be given a purely
static interpretation{}\cite{Obu80}\negthinspace \cite{CS80} and it formally
coincides with Reggeon field theory.{}\cite{BD74} This field theory contains
nonlinearities of lowest (=cubic) order, and the general expectation is that
it thus describes various processes with an active state for a one component
order parameter. Essential preconditions are that the particles diffuse and
that there are no conservation laws.

Recently T\"{a}uber et al.{}\cite{THH97} introduced a model describing DP
processes for $N$ particle types $A$, $B$, ... coupled by transmutation
reactions. For $2$ particle species the model may be defined by the reaction
equations 
\begin{eqnarray}
A &\rightarrow &2A,\;\;\;A\rightarrow 0,\;\;\;2A\rightarrow 0,\;\;\;A_{%
\mathbf{n}}\stackrel{\lambda }{\rightarrow }A_{\mathbf{n}+\mathbf{e}},
\label{Reactions} \\
B &\rightarrow &2B,\;\;\;B\rightarrow 0,\;\;\;2B\rightarrow 0,\;\;\;B_{%
\mathbf{n}}\stackrel{\lambda }{\rightarrow }B_{\mathbf{n}+\mathbf{e}}, 
\nonumber \\
&&A\stackrel{\sigma }{\rightarrow }B.  \nonumber
\end{eqnarray}
The particles may branch, decay and annihilate with some given rates.
Together with the diffusion reaction (rate $\lambda $) this yields normal
directed percolation. The two DP processes are coupled by the transmutation
reaction (rate $\sigma $). An interesting interpretation of the reactions (%
\ref{Reactions}) arises when one considers the particle species as particles
in different layers on a substrate. The reactions (\ref{Reactions}) then are
a model for the adsorption and desorption of particles.{}\cite{THH97}

A field theory corresponding to the process (\ref{Reactions}) may be derived
by standard techniques\cite{DOI76}\allowbreak \cite{GS80}\allowbreak \cite
{MG98}, the action functional reads{}\cite{THH97} 
\begin{eqnarray}
S &=&\int d^{d}xdt\left\{ -\widetilde{\phi }\left[ \partial
_{t}+r_{0}^{\left( \phi \right) }-\nabla ^{2}\right] \phi -\widetilde{\psi }%
\left[ \partial _{t}+r_{0}^{\left( \psi \right) }-\nabla ^{2}\right] \psi
+\sigma \psi \widetilde{\phi }\right\}  \label{Action} \\
&&+\int d^{d}xdt\left\{ \frac{g_{\psi }}{2\sqrt{K_{d}}}\psi \widetilde{\psi }%
^{2}-\frac{g_{\varphi }}{2\sqrt{K_{d}}}\phi ^{2}\widetilde{\phi }\right\}
+\int d^{d}xdt\left\{ \frac{u_{\varphi }}{2\sqrt{K_{d}}}\phi \widetilde{\phi 
}^{2}-\frac{u_{\psi }}{2\sqrt{K_{d}}}\psi ^{2}\widetilde{\psi }\right\} . 
\nonumber
\end{eqnarray}
Here $\phi $ corresponds to $A$, $\psi $ corresponds to $B$, and quartic
terms that are irrelevant in the renormalization group sense have been
dropped. The constant $K_{d}$ is ${2^{1-d}\pi ^{-d/2}/\Gamma \left( {d/2}%
\right) }$. To reach the critical point the parameters $r_{0}^{\left( \phi
\right) }$ and $r_{0}^{\left( \psi \right) }$ (or their renormalized
counterparts) have to be adjusted to $0$. Here we shall only consider the
cases $r_{0}^{\left( \phi \right) }=r_{0}^{\left( \psi \right) }$ or $%
r_{0}^{\left( \phi \right) }=0$.

For a transmutation rate $\sigma =0$ $S$ decomposes into uncoupled $A$ and $%
B $ DP action functionals. In fact, it follows from the reaction equations (%
\ref{Reactions}) or the Langevin equations corresponding to $S$, that
response functions of the type $\left\langle \widetilde{\phi }^{m}\phi
^{n}\right\rangle $ and $\left\langle \widetilde{\psi }^{m}\psi
^{n}\right\rangle $ are $\sigma $-independent. For $A$ the reaction $%
A\rightarrow B$ has the same effect as $A\rightarrow 0$, and for $B$ $\sigma 
$ has no effect as long as there there are no $A$ particles. The coupling is
only visible in the response functions $\left\langle \widetilde{\psi }%
^{m}\phi ^{n}\right\rangle $. This implies that the critical exponents $%
\beta $, $\nu $ and $z$ retain their DP values.{}\cite{THH97} It is also of
interest to note that $g_{\phi }$ and $u_{\phi }$ may be given the same
value by rescaling $\phi $ or $\widetilde{\phi }$, making the action
invariant under the transformation $\left\{ \phi \longleftrightarrow -%
\widetilde{\phi },\quad t\longleftrightarrow -t\right\} $ in the $\sigma =0$
case. It is equally well possible to fix $g_{\phi }=1$ and to get a
renormalization group fixed point $g_{\phi }=O\left( 1\right) $ and $u_{\phi
}=O\left( 4-d\right) $ instead of $g_{\phi }=\sqrt{O\left( 4-d\right) }$ and 
$u_{\phi }=\sqrt{O\left( 4-d\right) }$, and analogeously for $g_{\psi }$ and 
$u_{\psi }.$

T\"{a}uber et al.{}\cite{THH97} examined the model (\ref{Action}) for $%
\sigma \neq 0$ and noticed that additional cubic interactions $\int
d^{d}xdt\left\{ s_{1}\psi \widetilde{\psi }\widetilde{\phi }+\frac{s_{2}}{2}%
\psi \widetilde{\phi }^{2}+\frac{s_{3}}{4}\psi ^{2}\widetilde{\phi }%
+s_{4}\phi \psi \widetilde{\phi }\right\} $ are generated in the loop
expansion and must be taken into account in a renormalization group
calculation. The complete model displays two fixed lines in the space of the
coupling constants, one of them stable. This result was confirmed and
generalized by Janssen{}\cite{Jan99} , who also showed that the parameter $%
\sigma $ carries the same scaling dimension as $r_{0}-r_{0c}$. At any rate, $%
\sigma $ is strongly relevant, and a quick crossover to a strong coupling
regime will occur.

Here we are interested in this strong coupling regime and now set $\sigma =1$%
. For $\sigma \neq 0$ there is a ``rapidity''{}symmetry{}\cite{Jan99} 
\begin{equation}
\phi \longleftrightarrow -\widetilde{\psi },\;\;\widetilde{\phi }%
\longleftrightarrow -\psi ,\;\;t\longleftrightarrow -t,\;\;g_{\phi
}\longleftrightarrow g_{\psi },\;\;u_{\phi }\longleftrightarrow u_{\psi }.
\label{RapiditySym}
\end{equation}
In a formal sense the term \emph{unidirectional coupling }thus is somwhat
misleading. To simplify the calculation we rescale the fields to get $%
g=g_{\phi }=g_{\psi }$.

\subsubsection*{Dimensional analysis}

The goal now is a $2$-loop renormalization group (RG) calculation for the
model (\ref{Action}) in the $\sigma =1$ case. A first and elementary step of
any RG calculation is a dimensional analysis of the action integral. From
this one obtains the critical dimension of the model and the canonical wave
vector dimensions of the fields, coordinates and coupling constants.

There are some choices here, and it is of interest to formalize the (simple)
procedure. The condition that the terms of the action integral be
dimensionless leads to linear equations for the canonical wave vector
dimensions of the fields and coordinates. Let us call the number of fields
plus the number of coordinates the \emph{model order} of the field theory (6
in the case of Eq.(\ref{Action})). The determination of the model order
minus $1$ canonical dimensions of the fields and coordinates and of one
coupling constant (or the critical dimension instead) requires at least
model order linearly independent terms from the action integral. For $%
\varepsilon =0$ the linear equations may also be interpreted as equations
for a hyperplane in a model order dimensional exponent space. The integer
valued coordinates of this space are the exponents occurring in the terms of
the action integral, the canonical wave vector dimensions are the components
of the normal vector of this hyperplane. Selecting the terms of a critical
model thus amounts to determining a hyperplane in a model order\emph{\ }%
dimensional space. The normal vector of this hyperplane ($1$, $2$; $2$, $2$, 
$4$, $4$) in the case of Eq.(\ref{Action})) together with the critical
dimension, is a good signature for a critical model.\footnote{%
The normal vector of the critical hyperplane for $\sigma =0$ is ($1$, $2$; $%
2 $, $2$, $2$, $2$).} The terms in the halvespace containing the origin are
relevant, the terms in the other halfspace are irrelevant. The procedure of
selecting a critical hyperplane may also be captured in a computer
program.{} \cite{Kanon03}

For the action integral (\ref{Action}) the result for a hyperplane
containing the $\psi \widetilde{\varphi }$-term is a critical dimension $%
d_{c}=6$, and with $\varepsilon =d_{c}-d$, 
\[
\left[ g\right] =\varepsilon /2,\;\;\left[ u_{i}\right] =-2+\varepsilon
/2,\;\;\;\;\left[ \omega \right] =2,\;\;\;\;\left[ \phi \right] =\left[ 
\widetilde{\psi }\right] =2-\varepsilon /2,\;\;\;\;\left[ \widetilde{\phi }%
\right] =\left[ \psi \right] =4-\varepsilon /2. 
\]
Since it is crucial for the consistency of the model it must be checked now 
\begin{table}[tbp]
\begin{tabular}{|c|c|c|c|c|c|}
\hline
& $[\phi]=2$ & $[\widetilde{\psi }]=2$ & $[\widetilde{\phi }]=4$ & $[\psi]=4$
& Dimension \\ \hline
a) & 1 & 1 & * & * & $\geq 12$ \\ 
b) & 1 & 0 & * & 0 & $\geq 6$ \\ 
c) & 2 & 0 & * & 0 & $\geq 8$ \\ 
d) & 0 & 1 & 0 & * & $\geq 6$ \\ 
e) & 0 & 2 & 0 & * & $\geq 8$ \\ 
f) & 0 & 0 & 1 & 1 & $\geq 8$ \\ \hline
\end{tabular}
\caption{Field exponents and dimensions of possibly relevant terms (for the $%
u_{i}=0$ model). The asterisk denotes an exponent $\geq 1$.}
\label{Tab_TermExponents}
\end{table}
that the action (\ref{Action}) with $u_{i}=0$ and without the $s_{i}$ terms
contains all terms relevant or marginal near $d=6$, that may be generated
from other terms present originally, even irrelevant ones. The essential
fact to note to this purpose is that any vertex with an external $\phi $ at
least also contains one external $\widetilde{\phi }$, and vice versa for $%
\psi $: an external $\widetilde{\psi }$ implies at least one external $\psi $%
. This comes about because in the direction of increasing time a $\phi $
line may be converted to $\widetilde{\phi }$, but it cannot disappear. And
in the direction of decreasing time a $\widetilde{\psi }$ line may be
converted to $\psi $, but cannot disappear (see also Fig.(\ref{Fig_2loop})).
In Table (\ref{Tab_TermExponents}) we have listed the $\phi $ and $%
\widetilde{\psi }$ exponents of potentially dangerous terms for combinations
of $\phi $ and $\widetilde{\psi }$ exponents. The consistency of the action (%
\ref{Action}) follows from the fact that the canonical dimensions of the
field monomials are $\geq 8\equiv d_{c}+\left[ \omega \right] $, with the
exception of the $r_{0}$ terms that must be adjusted to $0$ anyway.

\subsubsection*{RG calculation}

\begin{figure}[tbp]
\includegraphics[clip]{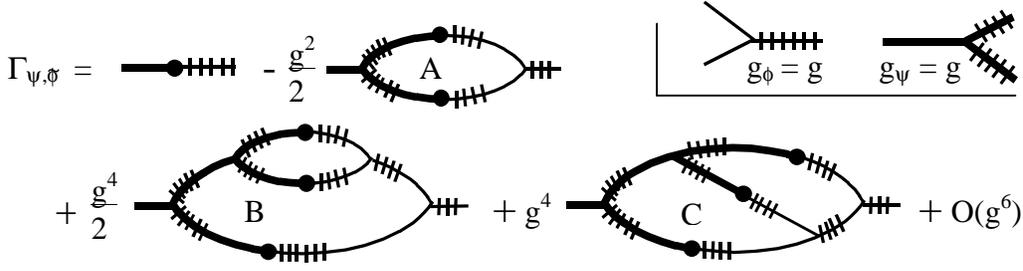}
\caption{Contributions to the $\Gamma _{\protect\sigma }$ vertex up to 2
loop order. On the top right there are also shown the graphs for the
interactions. The heavy line symbolizes the $\protect\psi $, the normal line
the $\protect\phi $ field. The dot denotes the $\Gamma _{\protect\sigma }$
vertex in tree approximation. }
\label{Fig_2loop}
\end{figure}

It follows from the dimensional analysis that the $u_{i}$ couplings are
strongly irrelevant and may be dropped, at least close to $d=6$. The Feynman
graphs for a $2$-loop RG calculation for the simplified action (\ref{Action}%
) are depicted in Fig.(\ref{Fig_2loop}). It is only the $\Gamma _{\sigma }$
vertex 
that gets renormalized. To generate a contribution to the $g_{\phi }$
interaction or to the $\phi $ propagator would require an irrelevant $\phi 
\widetilde{\phi }^{m}\widetilde{\psi }^{n}$ vertex in the left corner of the
time ordered diagram, and analogeously for the $g_{\psi }$ interaction
(rapidity inversion symmetry). Consequently only one independent $Z$ factor $%
Z_{\sigma }=\left( Z_{\psi }Z_{\widetilde{\phi }}\right) ^{1/2}$ is
required. For the individual fields this leads to 
\[
Z_{\psi }=Z_{\widetilde{\phi }}=Z_{\sigma },\;\;\;Z_{\widetilde{\psi }%
}=Z_{\phi }=Z_{\sigma }^{-1}. 
\]
The renormalized vertex function $\Gamma _{\sigma }$ then is 
\begin{equation}
\Gamma _{\sigma }^{\left( R\right) }\left( k^{2},\omega ,\tau ,g_{R},\mu
\right) =Z_{\sigma }\Gamma _{\sigma }\left( k^{2},\omega ,\tau ,g\right) ,
\label{Renormalization}
\end{equation}
where $\tau =r_{0}^{\left( \psi \right) }-r_{0c}^{\left( \psi \right)
}\equiv \tau _{R}$ and $g_{R}=Z_{\sigma }^{-1/2}g\mu ^{-\varepsilon /2}$ is
the dimensionless renormalized coupling constant. The condition that $g_{R}$
reaches a fixed point asymtotically\ requires $Z_{\sigma }\thicksim \mu
^{-\varepsilon }$.

The evaluation of the graphs of Fig.(\ref{Fig_2loop}) with standard
techniques{}\cite{BD74} yields 
\begin{eqnarray*}
A &=&\mu ^{-\varepsilon }\left( \frac{1}{4\varepsilon }+\frac{1}{16}\ln
\left( 4\right) -\frac{3}{16}+O\left( \varepsilon \right) \right) , \\
B &=&\mu ^{-2\varepsilon }\left( \frac{1}{32\varepsilon ^{2}}+\frac{12\ln
\left( 4\right) -6\ln \left( 3\right) -23}{384\varepsilon }+O\left( 1\right)
\right) , \\
C &=&\mu ^{-2\varepsilon }\left( \frac{1}{96\varepsilon }+\frac{\ln \left(
4\right) -\ln \left( 3\right) }{32\varepsilon }+O\left( 1\right) \right) .
\end{eqnarray*}
Using an external frequency $-i\omega =\mu ^{2}$ and minimal subtraction we
then find the stable RG fixed point

\begin{equation}
\left( g_{R}^{2}\right) _{MinSub}^{\ast }=8\varepsilon +\left( 4+40\ln
\left( 4/3\right) \right) \varepsilon ^{2}+O\left( \varepsilon ^{3}\right) .
\label{FixedPoint}
\end{equation}
The $\varepsilon $-expansion (\ref{FixedPoint}) does not seem to converge
for $\varepsilon =1$, and it seems to be impossible to extract a numerical
value for $g_{R}^{2\ast }$ for higher $\varepsilon $ values. However, the
critical exponents may be given in closed form. From Eq.(\ref
{Renormalization}) and the $\mu $-independence of the unrenormalized vertex
function $\Gamma _{\sigma }$ it follows 
\begin{equation}
\Gamma _{\sigma }^{\left( R\right) }\left( \frac{k}{\mu _{0}},\frac{\omega }{%
\mu _{0}^{2}},\frac{\tau _{R}}{\mu _{0}^{2}},g_{R}\left( \mu _{0}\right)
\right) =Z_{\sigma }^{-1}\left( \mu \right) \Gamma _{\sigma }^{\left(
R\right) }\left( \frac{k}{\mu },\frac{\omega }{\mu ^{2}},\frac{\tau _{R}}{%
\mu ^{2}},g_{R}\left( \mu \right) \right) \thicksim k^{\varepsilon }.
\label{ScalingGammaSigma}
\end{equation}
This finally leads to the scaling equivalence relations for the response
function, 
\[
\left\langle \widetilde{\psi }\phi \right\rangle \left( k,\omega ,\tau
\right) =\left\langle \widetilde{\psi }\psi \right\rangle \Gamma _{\sigma
}\left\langle \widetilde{\phi }\phi \right\rangle \thicksim
1/k^{4-\varepsilon }\thicksim 1/\omega ^{2-\varepsilon /2}\thicksim 1/\tau
^{2-\varepsilon /2}. 
\]
This exact result is valid f\"{u}r $d>4$ and in the inactive phase $\tau
\geq 0$.

It was argued above that the response functions $\left\langle \widetilde{%
\phi }^{m}\phi ^{n}\right\rangle $ and $\left\langle \widetilde{\psi }%
^{m}\psi ^{n}\right\rangle $ are independent of $\sigma $, and one might
think that it is inconsistent with this normal DP behaviour to have
nontrivial $Z$ factors and anomalous dimensions for $\phi $, $\widetilde{%
\phi }$, $\psi $ and $\widetilde{\psi }$. Actually this is not the case,
because general response functions of this type cannot be generated from the
action with $u_{i}=0$.

\subsubsection*{Summary}

It was shown that the problem of unidirectionally coupled directed
percolation may be solved with RG techniques above $d=4$ by expanding around 
$d_{c}=6$. In the inactive phase ($\tau \geq 0$) the scaling dimensions may
be given exactly.

In principle the solution $g=O\left( \varepsilon \right) $, $u_{i}=0$ found
for $4<d<d_{c}$ may extended contineously to $d<4$ with $g=O\left( \left(
4-d\right) ^{0}\right) $ and $u_{i}=O\left( 4-d\right) $. In this way the $%
\left\langle \widetilde{\phi }^{m}\phi ^{n}\right\rangle $ and $\left\langle 
\widetilde{\psi }^{m}\psi ^{n}\right\rangle $ response functions can retain
their normal DP behaviour. This also agrees with the fact that the scaling
relations $\phi \thicksim \widetilde{\psi }\thicksim k^{2}$ and $\widetilde{%
\phi }\thicksim \psi \thicksim k^{4-\varepsilon }$ coincide with the
uncoupled DP scaling relations at $d=4$.

The situation below $d=4$ is more complicated for the $\left\langle 
\widetilde{\psi }\phi \right\rangle $ response function, because here also
the $u_{i}$ interactions and possibly the mixed interactions considered by
T\"{a}uber et al.{}\cite{THH97} get relevant. T\"{a}uber et al. derive a
critical exponent $\beta _{2}=1/2-(4-d)/8+O\left( 4-d\right) ^{2}$
describing the expectation value of $\phi $ in the active phase in an
expansion around $d=4$. This procedure was criticized by Janssen{}\cite
{Jan99} in that it relies on the assymption that an exponentiation of
logarithms were possible. At any rate, for a finite coupling $\sigma $ the
diagrams of the type shown in Fig.(\ref{Fig_2loop}) must be taken into
account - all of them are IR singular already below $d_{c}=6$. The situation
also gets more involved in the active phase, where the irrelevant coupling
constant $u_{\psi }$ plays an important role, limiting the growth of $\psi $%
: $u_{\psi }$ is a ``dangerous irrelevant parameter'' above $d=4$, that
becomes a normal relevant parameter below $d=4$.

That the $1$-loop diagram ``$A$'' of Fig.(\ref{Fig_2loop}) poses a problem
for an RG calculation and an expansion around $d=4$ was already noticed by
Goldschmidt{}\cite{Gol98} (diagram ``$c$'' of this letter) and T\"{a}uber et
al.\cite{THH97}. An inspection of the diagrams for the model with $u_{i}=0$
indeed shows that $\beta _{2}$ is non-classical ($\neq 1/2$) already below $%
d=6$.

There is some similarity with the critical dynamics of the Heisenberg
ferromagnet (model $J$){}\cite{MM75}, where the dynamics is nonclassical
below $d=6$, while the statics is nonclassical below $d=4$. Likewise the
static critical exponents of model $J$ are independent of the dynamics. An
essential difference is that the model $J$ dynamics adds new coordinates and
fields, leaving room for a new critical exponent. In contrast, the scaling
dimensions of the coupled DP problem are completely determined by normal DP,
and one would expect $\beta _{2}=\beta $.

A dimensional analysis for the general coupled DP problem with $n$ particle
species (or $n$ layers) shows that here also only two nonlinearities of the
type displayed in Fig.(\ref{Fig_2loop}) are relevant near the critical
dimension, which now is $d_{c}\left( n\right) =2\left( n+1\right) $. These
nonlinearities act in the first and the last layer, and below $d_{c}\left(
n\right) $ more and more other nonlinearities get relevant. This indicates
that the RG in the form used here may not be the appropriate technique to
solve the general case. As a first step it would be of interest to fully
understand the $n=2$ case in the active phase and below $d=4$.


\medskip

\begin{thebibliography}{99}
\bibitem{Obu80}  S.P. Obukhov, Physica \textbf{110}A (1980) 145.

\bibitem{CS80}  J. L. Cardy and R. L. Sugar, J. Phys. A\textbf{13} (1980)
L423.

\bibitem{Sch72}  F. Schl\"{o}gl, Z. Phys. \textbf{253} (1972) 147.

\bibitem{BD74}  J.B. Bronzan and J.W. Dash, Phys. Rev. D\textbf{10} (1974)
4208, Phys. Rev. D\textbf{12} (1975) 1850.

\bibitem{THH97}  U.C. T\"{a}uber, M.J. Howard, and H. Hinrichsen, Phys. Rev.
Lett. \textbf{80} (1998) 2165; eprint: cond-mat/9709057. Phys. Rev. Lett. 
\textbf{81} (1998) 2179.

\bibitem{DOI76}  M. Doi, Journal of Physics A \textbf{9} (1976) 1465,
Journal of Physics A \textbf{9} (1976) 1479.

\bibitem{GS80}  P. Grassberger and P. Scheunert, Fortschr. Physik. \textbf{28%
} (1980) 547.

\bibitem{MG98}  D.C. Mattis und M.L. Glasser, Rev. Mod. Phys. \textbf{70}
(1998) 979.

\bibitem{Jan99}  H.K. Janssen, eprint: cond-mat/9901188.

\bibitem{Kanon03}  Available under the URL ``http://www.simtel.net'' as
KANON.

\bibitem{Gol98}  Y.Y. Goldschmidt, Phys. Rev. Lett. \textbf{81} (1998) 2178.

\bibitem{MM75}  S.K. Ma and G.F. Mazenko, Phys. Rev. B\textbf{11} (1975)
4077.
\end{thebibliography}
\end{document}